\documentclass[twocolumn]{aastex631}
\usepackage{color}
\usepackage{graphicx}
\usepackage{amsmath}
\usepackage{amssymb}	
\usepackage{diagbox}
\usepackage{threeparttable}

\usepackage{ulem}
\usepackage{natbib}
\usepackage{float}
\usepackage{longtable, threeparttablex, booktabs, url}
\usepackage{verbatim}

\usepackage{enumitem}

\usepackage{bm}

\begin{document}
\correspondingauthor{Di Li; Pei Wang; Hua-Xi Chen}
\email{dili@mail.tsinghua.edu.cn;\\wangpei@nao.cas.cn;\\chenhuaxi@zhejianglab.com}

\author[0000-0001-9956-6298]{Jian-Hua Fang}
\affiliation{Research Center for Astronomical Computing, Zhejiang Laboratory, Hangzhou 311121, China}

\author[0000-0003-3010-7661]{Di Li}
\affiliation{Department of Astronomy, Tsinghua University, Beijing 100084, China}
\affiliation{National Astronomical Observatories, Chinese Academy of Sciences, Beijing 100101, China}
\affiliation{Research Center for Astronomical Computing, Zhejiang Laboratory, Hangzhou 311121, China}

\author[0000-0002-3386-7159]{Pei Wang}
\affiliation{National Astronomical Observatories, Chinese Academy of Sciences, Beijing 100101, China}
\affiliation{Institute for Frontiers in Astronomy and Astrophysics, Beijing Normal University, Beijing 102206, China}

\author[0009-0000-6108-2730]{Hua-Xi Chen}
\affiliation{Research Center for Astronomical Computing, Zhejiang Laboratory, Hangzhou 311121, China}

\author{Han Wang}
\affiliation{Research Center for Astronomical Computing, Zhejiang Laboratory, Hangzhou 311121, China}

\author[0000-0002-7420-9988]{Deng-Ke Zhou}
\affiliation{Research Center for Astronomical Computing, Zhejiang Laboratory, Hangzhou 311121, China}

\author{Qin-Ping Bao}
\affiliation{Research Center for Astronomical Computing, Zhejiang Laboratory, Hangzhou 311121, China}

\author{Hai-Yan Li}
\affiliation{Research Center for Astronomical Computing, Zhejiang Laboratory, Hangzhou 311121, China}

\author{Jing-Jing Hu}
\affiliation{Research Center for Astronomical Computing, Zhejiang Laboratory, Hangzhou 311121, China}

\author[0000-0001-5649-2591]{Jin-Tao Xie}
\affiliation{School of Computer Science and Engineering, Sichuan University of Science and Engineering, Yibin 644000, China}

\author{Xiao-Dong Ge}
\affiliation{Research Center for Astronomical Computing, Zhejiang Laboratory, Hangzhou 311121, China}

\author[0000-0002-0475-7479]{Yi Feng}
\affiliation{Research Center for Astronomical Computing, Zhejiang Laboratory, Hangzhou 311121, China}
\affiliation{Institute for Astronomy, School of Physics, Zhejiang University, Hangzhou 310027, China}

\author[0000-0003-4811-2581]{Dong-Hui Quan}
\affiliation{Research Center for Astronomical Computing, Zhejiang Laboratory, Hangzhou 311121, China}
\affiliation{Institute for Astronomy, School of Physics, Zhejiang University, Hangzhou 310027, China}

\author{Zhi-Xuan Kang}
\affiliation{Research Center for Astronomical Computing, Zhejiang Laboratory, Hangzhou 311121, China}

\author{Xue-Rong Guo}
\affiliation{Research Center for Astronomical Computing, Zhejiang Laboratory, Hangzhou 311121, China}

\author{Chen-Wu Jin}
\affiliation{Research Center for Astronomical Computing, Zhejiang Laboratory, Hangzhou 311121, China}

\author{Zhi-Lin Wang}
\affiliation{Alibaba Group, Hangzhou 311121, China}

\author[0000-0002-9579-6739]{Jia-Ying Xu}
\affiliation{Research Center for Astronomical Computing, Zhejiang Laboratory, Hangzhou 311121, China}

\author{Chen-Chen Miao}
\affiliation{Research Center for Astronomical Computing, Zhejiang Laboratory, Hangzhou 311121, China}

\author[0000-0002-1243-0476]{Ru-Shuang Zhao}
\affiliation{School of Physics and Electronic Science, Guizhou Normal University, Guiyang 550001, China}

\author[0000-0001-6651-7799]{Chen-Hui Niu}
\affiliation{College of Physical Science and Technology, Central China Normal University, Wuhan 430079, China}

\title{TransientVerse: A Comprehensive Real-Time Alert and Multi-Wavelength Analysis System for Transient Astronomical Events}

\begin{abstract}
Transient astrophysical events are characterized by short timescales, high energy, and multi-wavelength radiation, often accompanied by violent energy releases. These phenomena are a major focus of modern astronomical research. To reveal their underlying physical mechanisms, near-real-time, multi-wavelength, and multi-messenger follow-up observations are essential. However, current transient alert systems face multiple challenges, including fragmented messages, inconsistent formats, and difficulties in retrospective analysis, all of which hinder the efficiency of triggering observations. This paper presents \texttt{TransientVerse}, an innovative real-time database platform to integrate and disseminate transient alerts. The platform uses an automated pipeline to integrate real-time alerts from multiple sources (e.g., ATel, VOEvent, and GCN). It structures unstructured text data into a dual-format database for transient alerts by using open-source large language models. TransientVerse offers retrospective searches, data visualization, literature reviews, and customized subscriptions for efficient event tracking and analysis. Additionally, for Fast Radio Bursts (FRBs), the platform provides real-time statistics on repeat burst rates across different time intervals and alerts astronomers about high-frequency burst sources, enabling rapid follow-up observations and optimizing the use of limited observation windows. TransientVerse improves the efficiency of acquiring transient events in real time, lowers the technical barriers for simultaneous observations, and provides robust technical support for multi-wavelength, multi-messenger time-domain astronomy and astrophysics studies.
\end{abstract}

\keywords{Transient sources; Radio transient sources; High energy astrophysics; Time domain astronomy; Gamma-ray bursts}

\section{Introduction} \label{sec:intro}
Transient sources are astronomical events that suddenly appear and disappear over short timescales. 
These sources, often associated with relativistic compact objects, are typically linked to intense energy-release processes such as SuperNovae (SN), FRBs, and Gamma-Ray Bursts (GRBs). Due to the rapid and intense energy release, these sources exhibit extreme physical conditions, making them crucial for exploring related astrophysical phenomena. As a result, transient sources have become a central focus in modern astronomical research. The rapid development of multi-wavelength and multi-messenger astronomy has provided new perspectives for studying these transient astronomical events. 
Complementary cosmic messengers, including neutrinos, cosmic rays, and Gravitational Waves (GWs), enhance traditional electromagnetic observation methods, enriching our understanding of extreme astronomical events \citep{2017ApJ...848L..12A, 2019NatRP...1..585M}. 
Due to the short-duration nature of transient sources, observing these astronomical events imposes stringent requirements on both observational equipment and real-time alert systems. These systems are required to rapidly detect and track intense astronomical events, like SN and GRBs, and then share the discoveries with the global observational community, enabling other telescopes to carry out follow-up multi-wavelength and multi-messenger observations \citep{2016arXiv160603735S, 2019A&A...631A.147N, 2020APh...114...68A, 2021MNRAS.501.3272M, 2021ApJS..256....5R}.

With the construction and commissioning of next-generation astronomical facilities, multi-wavelength and multi-messenger observations have progressively emerged as a core research direction in modern astronomy. Traditional observations, focused on single electromagnetic bands (e.g., optical or radio) or individual messengers (such as neutrinos, cosmic rays, or GWs), can no longer comprehensively unveil the complex phenomena in the universe. Therefore, combining multi-wavelength and multi-messenger observations has become the inevitable choice for conducting a more comprehensive study of astrophysical processes.

In the electromagnetic spectrum, various large telescopes, such as the Square Kilometer Array (SKA; \citealt{2009IEEEP..97.1482D}), China Space Station Telescope (CSST; \citealt{2011SSPMA..41.1441Z, 2018MNRAS.480.2178C, 2019ApJ...883..203G}), Large Synoptic Survey Telescope (LSST; \citealt{2009arXiv0912.0201L, 2019ApJ...873..111I}), Advanced Telescope for High-Energy Astrophysics (ATHENA; \citealt{2018SPIE10699E..1GB}), and ground-based gamma-ray observatories such as Large High Altitude Air Shower Observatory (LHAASO; \citealt{2019arXiv190502773C}) and Cherenkov Telescope Array (CTA; \citealt{2011ExA....32..193A}), will cover a wide range of electromagnetic bands, from radio waves to optical, X-rays, and gamma-rays. These are expected to significantly enhance our observational capabilities for transient sources and related cosmic events. 

In the field of multi-messenger astronomy, high-energy neutrino telescopes, exemplified by the IceCube Neutrino Observatory (IceCube; \citealt{2013Sci...342E...1I}), cosmic ray detector arrays (such as LHAASO), and GW detectors (including LIGO, Virgo, and KAGRA; \citealt{2018LRR....21....3A}) supply complementary data for transient source research. Multi-messenger observations, in synergy with multi-wavelength electromagnetic ones, yield multidimensional perspectives of transient sources, unveiling more details about these phenomena. Simultaneously, these multi-messenger and multi-wavelength observations imply a large volume of real-time observational data and alert information.

In order to facilitate the real-time distribution of these alert messages to other astronomers for follow-up observations using multi-wavelength and multi-messenger instruments, it is essential to establish a workflow that can process and analyze these alert messages, thereby enabling their distribution, filtering, and classification. Although there are already many excellent astronomical alert platforms such as Astronomer's Telegram (Atel)\footnote{\url{https://astronomerstelegram.org/}}, Gamma-ray Coordinates Network (GCN)\footnote{\url{https://gcn.nasa.gov/}}, and Virtual Observatory Event (VOEvent)\footnote{\url{https://www.chime-frb.ca/voevents}}, these alerts are often published across multiple platforms, leading to incomplete information dissemination. Furthermore, some alert platforms do not use the machine-readable standard data formats developed by the International Virtual Observatory Alliance (IVOA), but instead release observations in human-readable formats, which poses challenges to batch processing and automation. To address the challenges of scattered, inconsistent, and non-retroactive alert messages, it is crucial to establish a platform that can integrate and push transient source alerts in real-time. This will significantly improve observational efficiency, enhance the comprehensive understanding of these transient phenomena, and promote near-real-time multi-wavelength follow-up observations of transient sources.

To address the growing challenges of observational data and alerts, we develop a real-time notification platform called ``TransientVerse"\footnote{\url{https://transientverse.zero2x.org/}}. TransientVerse collects and integrates transient source alerts from multiple sources, such as Atel, GCN, and VOEvent, through an automated collection pipeline. By leveraging open-source large language models, the platform parses textual data into structured formats, which are promptly delivered to users via standardized templates, facilitating multi-wavelength and multi-messenger follow-up observations. Additionally, TransientVerse provides retrospective query, visualization, and literature research functionalities, allowing users to efficiently analyze the collected alerts.

The paper is structured as follows: Section \ref{sec:science} discusses the scientific motivations; Section \ref{sec:architecture} describes the platform's architecture and key features; and Section \ref{sec:conclusion} presents a summary and future outlook.

\section{Science Drivers}\label{sec:science} 
Various transient astrophysical events, such as FRBs, GRBs, and GWs, require multi-wavelength and multi-messenger observations to reveal their origins, physical mechanisms, and temporal evolution. The scientific motivation behind the TransientVerse platform lies in providing real-time alerts, analysis, and follow-up observations of these events, enabling coordinated efforts across multi-wavelength and multi-messenger domains.

\subsection{Fast Radio Bursts} 
FRBs are the brightest transient phenomena in the radio band, with durations typically ranging from microseconds to milliseconds. Since the discovery of the first FRB \citep{2007Sci...318..777L}, over 800 FRB sources have been identified \citep{2021ApJS..257...59C, 2021Natur.598..267L, 2022Natur.606..873N, 2022Sci...375.1266F, 2022RAA....22l4002Z, 2023Univ....9..330X}. However, their origins and physical mechanisms remain a mystery. The first multi-wavelength simultaneous observation of the FRB 20200428 event was observed from the magnetar SGR 1935+2154 in our galaxy, where the burst was observed in both the X-ray and radio bands \citep{2020ApJ...898L..29M}, significantly advancing our understanding of FRB origins and mechanisms. Notably, the occurrence of a ``burst forest" in magnetar suggests that the source may have entered an active phase, potentially generating FRBs similar to FRB 20200428, offering a trigger for further research. Due to the short duration of FRB bursts, they are often not directly observable during coordinated multi-wavelength campaigns. Therefore, follow-up observations of their afterglows are especially important.

\subsection{Gamma-Ray Bursts}
GRBs are the most powerful electromagnetic bursts in the universe \citep{2015PhR...561....1K}, with durations ranging from a few milliseconds to several hours \citep{2011Natur.480...72T, 2013ApJ...766...30G, 2014ApJ...781...13L}. Despite over 50 years of observations of GRBs \citep{1973ApJ...182L..85K}, many aspects related to their jet physics, particle acceleration, radiation processes, and radiation mechanisms remain mysterious \citep{2019pgrb.book.....Z}. Simultaneous observations across multiple wavelengths offer the potential to address these questions. In 2022, the Fermi Gamma-Ray Burst Monitor (GBM) and the Swift Burst Alert Telescope (BAT) detected the brightest GRB ever recorded, named GRB 221009A \citep{2022ATel15660....1T, 2022GCN.32636....1V, 2023ApJ...946L..24W}. Subsequently, its counterpart was observed in radio, optical, and X-ray bands \citep{2022GCN.32635....1K, 2022GCN.32636....1V, 2022GCN.32634....1L, 2022GCN.32653....1B}; additionally, very high-energy emission was detected, with the LHAASO observatory detecting photons up to 18 TeV \citep{2022GCN.32677....1H}. Many experimental groups later conducted follow-up observations of its afterglow across the radio, millimeter, optical, X-ray, and gamma-ray bands \citep{2023ApJ...946L..23L}, significantly enhancing our understanding of GRB explosion mechanisms. Therefore, real-time notification of GRB alert information is crucial for coordinating multi-wavelength observations of GRB explosions and afterglows using combined telescope facilities.

\subsection{Gravitational Waves}
GWs are ripples in spacetime predicted by general relativity, propagating at the speed of light. They are expected to originate from extremely massive, strongly accelerated celestial systems, including symmetric non-rotating systems as well as asymmetric collapses or mergers of massive objects, such as black holes or neutron stars. Since the first GW event, GW170817, was independently detected by LIGO and Virgo \citep{2017GCN.21587....1L}, an increasing number of GW events have been observed successfully. The electromagnetic counterparts of GW170817 were detected across multiple wavelengths, including radio, ultraviolet, optical, infrared, and X-ray bands. In the radio and X-ray bands, signals were not detected initially but were observed approximately 16 and 9 days after the event, respectively. In the optical band, signals were independently detected within 11 hours of the explosion by multiple teams. In the ultraviolet band, a signal was observed early on and faded within 48 hours, while the optical and infrared bands showed redshift evolution within about 10 days. No signals were detected in the neutrino or high-energy gamma-ray bands. These multi-wavelength and multi-messenger observations support the hypothesis that the GW170817 event was triggered by the merger of two neutron stars in NGC 4993 \citep{2017ApJ...848L..12A}.

FRBs and GRBs might be associated with GWs. These high-energy burst events are likely to originate from similar extreme celestial systems, such as black hole or neutron star mergers, making them high-priority targets for observational studies. Their study could provide crucial insights into understanding high-energy explosive phenomena.

The search for multi-messenger and electromagnetic counterparts of GWs is crucial for understanding the origin of GW events. However, due to the significant uncertainty in the sky localization of GWs, the search for multi-messenger and electromagnetic counterparts becomes extremely challenging and requires additional telescopes to observe the region identified by the GW detectors. Furthermore, as the signals in different wavelength bands appear at different time intervals, covering hours, days, and weeks in various parts of the electromagnetic spectrum, the timely notification of GW alerts is crucial for coordinating multi-wavelength follow-up observations, which will help reveal the mechanisms behind GW production.

\section{Overall Architecture of the Transient Source Platform}\label{sec:architecture}
\begin{figure*}[ht] 
    \centering
    \includegraphics[width=0.9\textwidth]{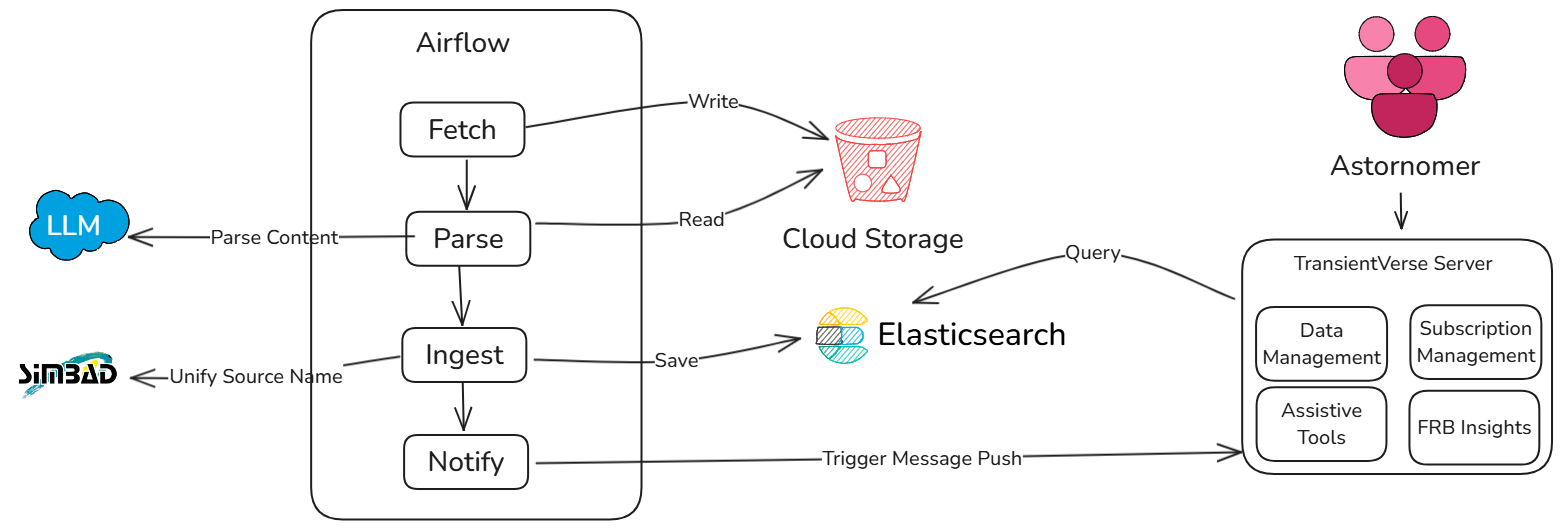}
    \caption{The architecture of the transient astronomical alert message collection and integration system. The system efficiently processes and manages alert messages through a data pipeline orchestrated by Airflow, which includes stages for fetching (Fetch), parsing (Parse), ingestion (Ingest), and notification (Notify). Parsed data is stored in Cloud Storage and indexed in Elasticsearch for quick querying. The TransientVerse server provides astronomers with tools for data management, subscription services, assistive utilities, and FRB insights, enabling streamlined monitoring and analysis of transient alert messages.} 
    \label{fig:system-architecture}
\end{figure*}

The architecture of the TransientVerse system is designed to integrate transient source alert messages, providing real-time notifications and enabling retrospective queries. It consists of two main components: the ingestion pipeline and application services. Figure \ref{fig:system-architecture} illustrates the role of each component in ensuring efficient data processing, storage, and user access.

\subsection{Ingestion Pipeline}
The ingestion pipeline, orchestrated by Airflow, is the core of TransientVerse, responsible for collecting, processing, and storing transient source alerts from multiple sources. It consists of three main stages:

\subsubsection{Alert Message Retrieval} 
The first step in the ingestion pipeline is alert message retrieval. It is primarily responsible for obtaining transient source alert messages from multiple alert sources and ensuring that incoming report files are stored securely and efficiently in different raw formats. To ensure high availability and durability of the data, we use cloud-based storage services, ensuring that all data can be reliably stored and accessed when needed. This design aims to streamline the process while maintaining stability, thereby ensuring the smooth execution of subsequent stages.

The platform currently supports the following key alert message sources:

\begin{itemize} 
    \item \textbf{Atel}: Atel is a web-based short-notice publication system designed for quickly reporting and commenting on new astronomical observations \citep{1998PASP..110..754R}. Its advantage lies in its simplicity and efficiency, allowing users to directly publish Telegram posts on the website, with the content being immediately made public, thus ensuring high real-time relevance. However, the content published by Atel lacks a standardized format and is presented in human-readable form, making it difficult to extract structured information (such as source names, telescope names, and coordinates) for machine automation. To address this, the platform will integrate open source large-language models to automate the extraction of this information.
    
    \item \textbf{VOEvent}: VOEvent, developed by IVOA, is a standard format for describing and transmitting information about transient astronomical events \citep{2014htu..conf..105S}. The VOEvent message source integrated into the platform is provided by the CHIME (Canadian Hydrogen Intensity Mapping Experiment) collaboration, which transmits real-time alerts about newly discovered and known FRBs \citep{2017arXiv171008155P}. These alert messages are distributed rapidly to observatories and astronomers worldwide and are automatically included in the Fast Radio Burst Catalog (FRBCAT) for real-time updates, making it easy for researchers to access and analyze related data \citep{2016PASA...33...45P}. The platform also tracks these repeated burst alerts and identifies FRBs with high frequency, pushing relevant information to users for follow-up observations of these frequent FRBs.

    \item \textbf{GCN}: GCN is a platform used for publishing information about GRBs, GW compact binary mergers, high-energy neutrino discoveries, and other astronomical events \citep{1995Ap&SS.231..235B, 2001AIPC..587..213B}. GCN provides two data formats: Circulars and Notices. 
    Circulars are readable, citable reports, often not in real-time, and include observations, quantitative short-term predictions, follow-up observation requests, or future observation plans \citep{2024AAS...24335913B}; Notices are real-time alerts automatically generated, typically involving event localization and detection information. Since Circulars lack a uniform message format and are presented in human-readable form, extracting structured information (such as source names, telescope names, and coordinates) is difficult for machine automation. The platform will use open-source large language model technologies to automate the extraction of this information, enabling structured processing for easier subsequent display and analysis. Notices contain a large amount of information about telescope localization and pointing, which has limited practical application for transient sources, so the platform will focus on filtering alerts directly related to transient source events to improve alert efficiency.
\end{itemize}

\begin{figure*}[ht]
    \centering
    \includegraphics[width=0.9\textwidth]{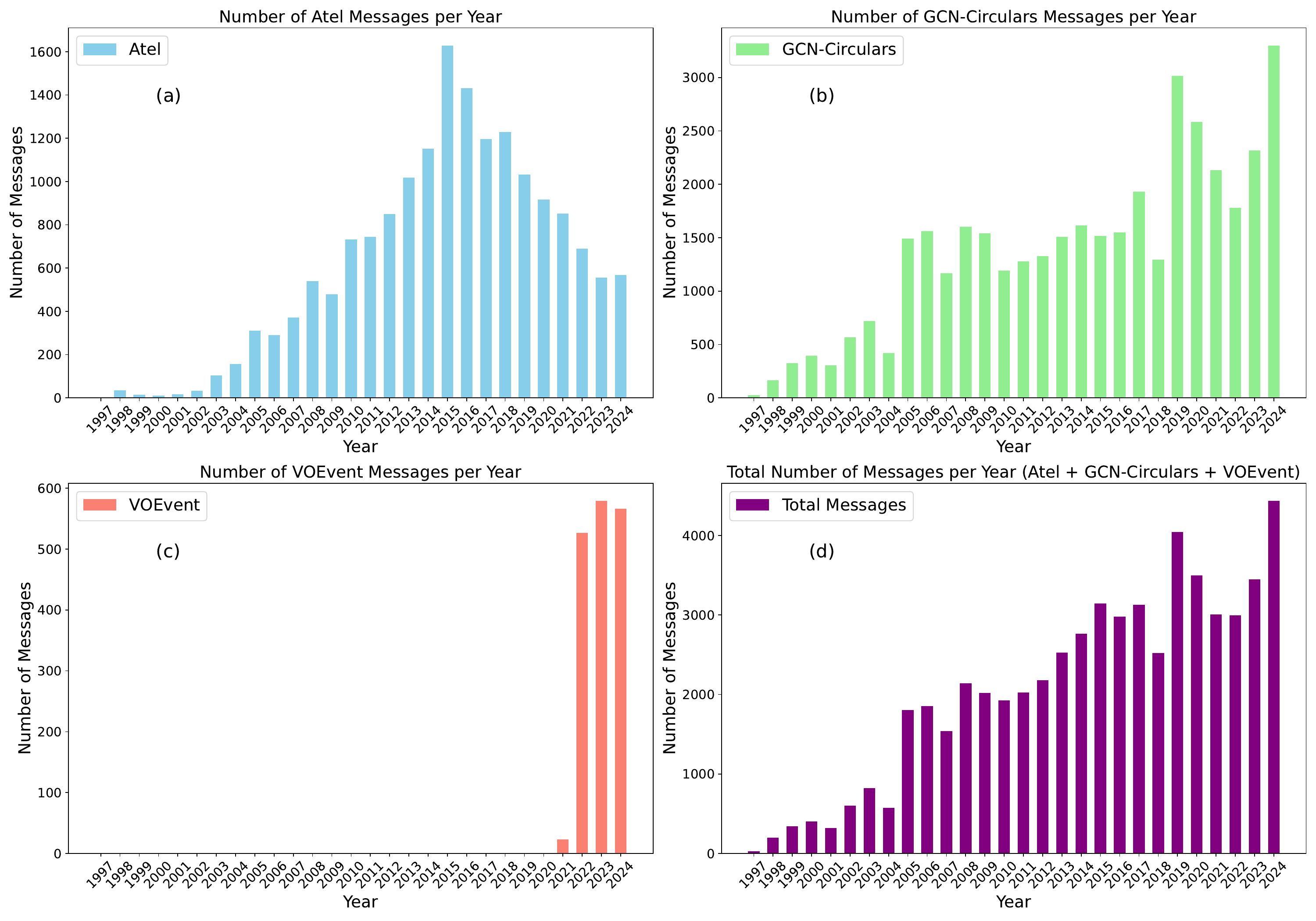}
    \caption{The annual number of messages from three astronomical sources: Atel (panel a), GCN-Circulars (panel b), and VOEvent (panel c). Panel d illustrates the combined message count from all three sources.}  
    \label{fig:combined_histogram_2x2}
\end{figure*}

In the platform's design, the acquisition and processing of alert messages are among the core components. Figure \ref{fig:combined_histogram_2x2} illustrates the number of alerts obtained each year from the Atel, GCN Circulars, and VOEvent sources. The first three subplots in the figure show the annual alert counts for each message source, while the last subplot presents the total alert count from all three sources for comparison, providing an overview of the data distribution across different sources. Additionally, Table \ref{tab:source_introduction} summarizes the characteristics, descriptions, and message formats of these message sources.

\begin{nolinenumbers}
\begin{table*}[htbp]
    \centering
    \caption{Description and Formats of Different Message Sources}
    \begin{tabular}{lp{10cm}l}
    \hline
    \textbf{Message Source\textsuperscript{a}} & \textbf{Description\textsuperscript{b}} & \textbf{Message Format\textsuperscript{c}} \\
    \hline
    \textbf{Atel} & A web-based short-notice publishing system that quickly reports new astronomical observations, with content publicly available in real-time. Simple and efficient. & Human-readable \\
    \hline
    \textbf{VOEvent} & A standard format developed by IVOA to describe and transmit transient astronomical event information, ensuring unified information sharing and processing. & Machine-readable \\
    \hline
    \textbf{Circulars} & Readable and citable quick bulletins, usually not in real-time, containing observational results, short-term predictions, follow-up observation requests, or future observation plans. & Human-readable \\
    \hline
    \textbf{Notices} & Automatically generated real-time alerts in machine-to-machine format, containing localization and detection information for astronomical events like GRBs and GWs. & Machine-readable \\
    \hline
    \end{tabular}
    \label{tab:source_introduction}
    \vspace{0.2cm}
    \raggedright
    
    \small{
        \hspace{1cm}\textsuperscript{a} \textbf{Message Source}: The name of the message source. \\
        \hspace{1cm}\textsuperscript{b} \textbf{Description}: A brief explanation of the type of information or service the source provides. \\
        \hspace{1cm}\textsuperscript{c} \textbf{Message Format}: The type of format in which the message is presented (human-readable or machine-readable).
    }
\end{table*}
\end{nolinenumbers}

\subsubsection{Report Parsing}
A standardized parsing procedure follows the report acquisition process to extract relevant information from the reports. This stage integrates Large Language Models (LLMs) with custom parsers to ensure high accuracy. The LLMs are fine-tuned using carefully designed prompts, while custom parsers handle specific file formats. Once the data is parsed, it is converted into a unified structure, ensuring that all extracted information follows a standardized format for efficient querying and analysis.

To optimize the performance of the LLMs in this text parsing task, we meticulously design prompts that include key components: role specification, task description, main objectives (with particular emphasis on fields like telescope names), and a well-crafted one-shot learning example. The role specification clarifies the model's function, guiding it to identify and extract structured data from scientific reports. The task description explicitly outlines the objectives, such as extracting transient names, telescope identifiers, and astronomical coordinates (RA and Dec). Special attention is given to handling variations in the format of critical fields like telescope names, ensuring the model can correctly interpret and process diverse naming conventions.

A parsing example is included in the prompt to provide a clear demonstration of how the model should interpret the text. This one-shot learning approach allows the LLM to learn from a single example and apply that knowledge across the entire report. The input-output format is also clearly defined, ensuring that the model’s responses are structured and consistent.

In this process, several core fields are selected for extraction, including Transient Name, Transient Type, Telescope, RA (Right Ascension), and Dec (Declination). The effectiveness of different models is evaluated based on accuracy, precision, and edit distance metrics. Through extensive testing and comparison, GPT-4 Turbo \citep{openai_gpt4_reference} and ERNIE-Bot 4.0 \citep{ernie_bot4_blog} were identified as the top-performing models for this task. These models demonstrated exceptional capabilities in parsing complex report formats and extracting the necessary data, offering a reliable solution that ensures consistent data quality for downstream analysis.

By combining fine-tuned LLM prompts with custom parsing techniques and a unified data structure, this approach significantly enhances the accuracy and efficiency of extracting structured data, facilitating seamless integration into subsequent analysis stages.

\subsubsection{Extraction to Search Engine}
Both raw and processed data are ingested into Elasticsearch (ES)\footnote{\url{https://www.elastic.co/cn/elasticsearch}}, a robust search and analytics engine capable of handling large datasets and providing real-time search capabilities.

To ensure consistency and accuracy, we integrate data from SIMBAD to retrieve coordinate positions and standardize transient names based on the source name \citep{2000A&AS..143....9W}. This process produces uniform fields, enabling efficient relational queries and linking follow-up events associated with the same transient source. Such standardization enhances data usability and simplifies the process of tracking subsequent observations.

Additionally, ES's geospatial capabilities are leveraged by mapping astronomical coordinates (RA and Dec) to a geographic coordinate system. This transformation allows the system to perform approximate geographic searches based on sky regions, significantly improving its ability to deliver relevant and precise results.

By combining standardized fields from SIMBAD with ES’s geospatial search features, the component provides a powerful and user-friendly solution for processing and querying astronomical data. Users can efficiently explore datasets and identify related observations.

\subsection{Application Services}

The application services layer is designed to provide a smooth user experience, making it easier to access and subscribe to transient source alerts. It offers the following key features: search, display, and subscription of transient source reports, as well as quick investigation of sources in the reports. Additionally, the platform includes tools for unit conversion and a ``Research Assistant" tool. Furthermore, for VOEvent FRB alerts, the platform uses historical alert data to highlight high-frequency FRB sources on sky maps, which are then pushed to astronomers. This feature facilitates follow-up observations of these high-frequency FRB events.

\subsubsection{Search for Transient Source Alert Messages}
\begin{figure*}[ht] 
    \centering
    \includegraphics[width=0.98\textwidth]{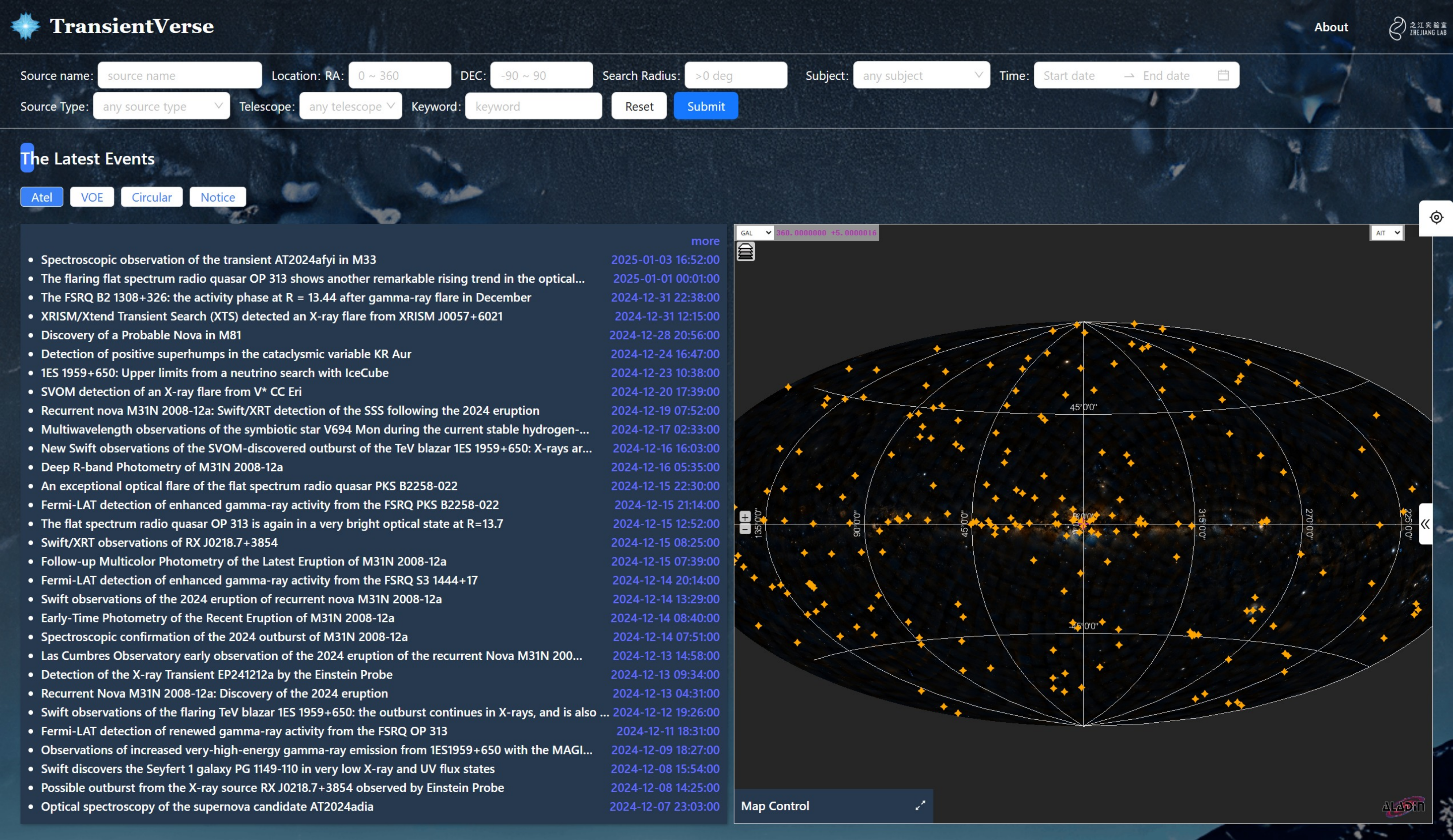}
    \caption{Home page of the TransientVerse platform. The top section provides a search interface with filters for source name, coordinates (with radius), subjects, time, source type, telescope, and keywords. The bottom section displays recent alerts and a skymap showing the spatial distribution of transient sources.} 
    \label{fig:transient_main}
\end{figure*}

The TransientVerse homepage, shown in Figure \ref{fig:transient_main}, features a search section at the top with fields for source name, coordinates (with angular radius), subject, time range, transient source type, observing telescope, and keywords. These criteria allow astronomers to filter messages and locate relevant transient source alerts. The bottom part of the page displays recent alerts along with a sky map showing the spatial distribution of the involved sources.

\begin{figure*}[ht] 
    \centering
    \includegraphics[width=0.98\textwidth]{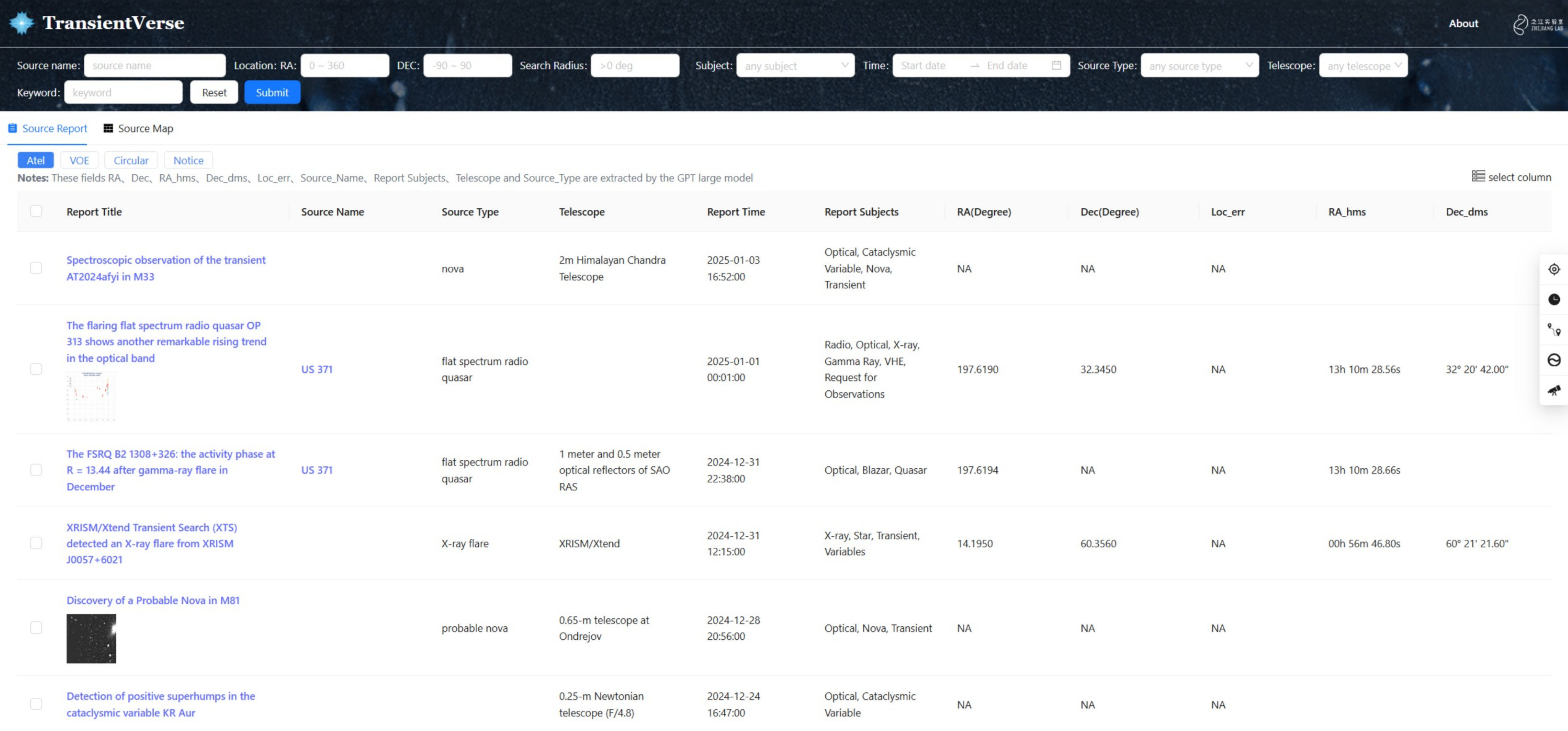}
    \caption{Structured alert data representation on TransientVerse. The \textbf{Report Title} column links to the original textual report, while the \textbf{Source Name} column provides detailed source information, including basic data, related alerts, and papers retrieved from Astrophysics Data System (ADS)\footnote{\url{https://ui.adsabs.harvard.edu/}}. Keywords such as \textbf{RA}, \textbf{Dec}, \textbf{RA\_hms}, \textbf{Dec\_hms}, \textbf{Loc\_err}, \textbf{Source Name}, \textbf{Report Subjects}, \textbf{Telescope}, and \textbf{Source Type} are extracted from the textual messages using LLMs.}
    \label{fig:transient_structed}
\end{figure*}

When a search is executed, structured transient source alert messages matching the criteria are shown, as seen in Figure \ref{fig:transient_structed}. If no criteria are specified, submitting the search form or clicking ``more" will lead to the latest structured alert messages. These messages are extracted through custom parsers and open-source LLMs.

\begin{table*}[ht]
    \centering
    \caption{Structured Data Fields from Various Alert Message Sources}
    \begin{tabular}{llcccc}
    \toprule
    \textbf{Field Name}$^{\ast}$ & \textbf{Description} & \textbf{Atel} & \textbf{VOEvent} & \textbf{Circular} & \textbf{Notice} \\
    \midrule
    \textbf{Report Title} & Title of the alert report & \checkmark & \checkmark & \checkmark & \checkmark \\
    \textbf{Source Name} & Name of the source & \checkmark & \checkmark & \checkmark & \checkmark \\
    \textbf{Source Type} & Type of the source & \checkmark & \checkmark & \checkmark & \checkmark \\
    \textbf{Telescope} & Name of the telescope & \checkmark & \checkmark & \checkmark & \checkmark \\
    \textbf{Report Time} & Time of the report publication & \checkmark & \checkmark & \checkmark & \checkmark \\
    \textbf{Report Subjects} & Classification of the report's subjects & \checkmark & \checkmark & \checkmark & \checkmark \\
    \textbf{RA (Degree)} & Right Ascension of the source (in degrees) & \checkmark & \checkmark & \checkmark & \checkmark \\
    \textbf{Dec (Degree)} & Declination of the source (in degrees) & \checkmark & \checkmark & \checkmark & \checkmark \\
    \textbf{Loc\_err} & Localization error (in degrees) & \checkmark & \checkmark & \checkmark & \checkmark \\
    \textbf{RA\_hms} & Right Ascension of the source (in HMS format) & \checkmark & \checkmark & \checkmark & \checkmark \\
    \textbf{Dec\_dms} & Declination of the source (in DMS format) & \checkmark & \checkmark & \checkmark & \checkmark \\
    \textbf{Event Type} & Type of the event & & \checkmark & & \\
    \textbf{Event Number} & Event number & & \checkmark & & \\
    \textbf{DM} & Dispersion Measure  of the event (in $\text{pc} \cdot \text{cm}^{-3}$) & & \checkmark & & \\
    \textbf{DM\_err} & Error of the Dispersion Measure (in $\text{pc} \cdot \text{cm}^{-3}$) & & \checkmark & & \\
    \textbf{DM\_ne} & Upper limit of Dispersion Measure (in $\text{pc} \cdot \text{cm}^{-3}$) & & \checkmark & & \\
    \textbf{DM\_ymw} & Time variation of the Dispersion Measure (in $\text{pc} \cdot \text{cm}^{-3}$) & & \checkmark & & \\
    \textbf{SNR} & Signal-to-Noise Ratio & & \checkmark & & \\
    \textbf{GAL\_LONG} & Galactic longitude of the source (in degrees) & & & & \checkmark \\
    \textbf{GAL\_LAT} & Galactic latitude of the source (in degrees) & & & & \checkmark \\
    \textbf{ECL\_LONG} & Ecliptic longitude of the source (in degrees) & & & & \checkmark \\
    \textbf{ECL\_LAT} & Ecliptic latitude of the source (in degrees) & & & & \checkmark \\
    \bottomrule
    \end{tabular}\\
    \parbox{\textwidth}{
        \raggedright
        $^{\ast}$ Field Name refers to the structured keywords extracted from the alert reports, summarizing the key metadata of astronomical events. \\
        $\checkmark$ indicates the presence of the corresponding field in the message source.
    }
    \label{tab:tab2}
\end{table*}

\subsubsection{Display of Transient Source Alerts}
TransientVerse processes raw textual data into a structured format using open-source LLMs. It offers three main display methods for transient source alerts: textual data, structured data, and interactive sky map.

\begin{figure*}[ht]
    \centering
    \includegraphics[width=0.98\textwidth]{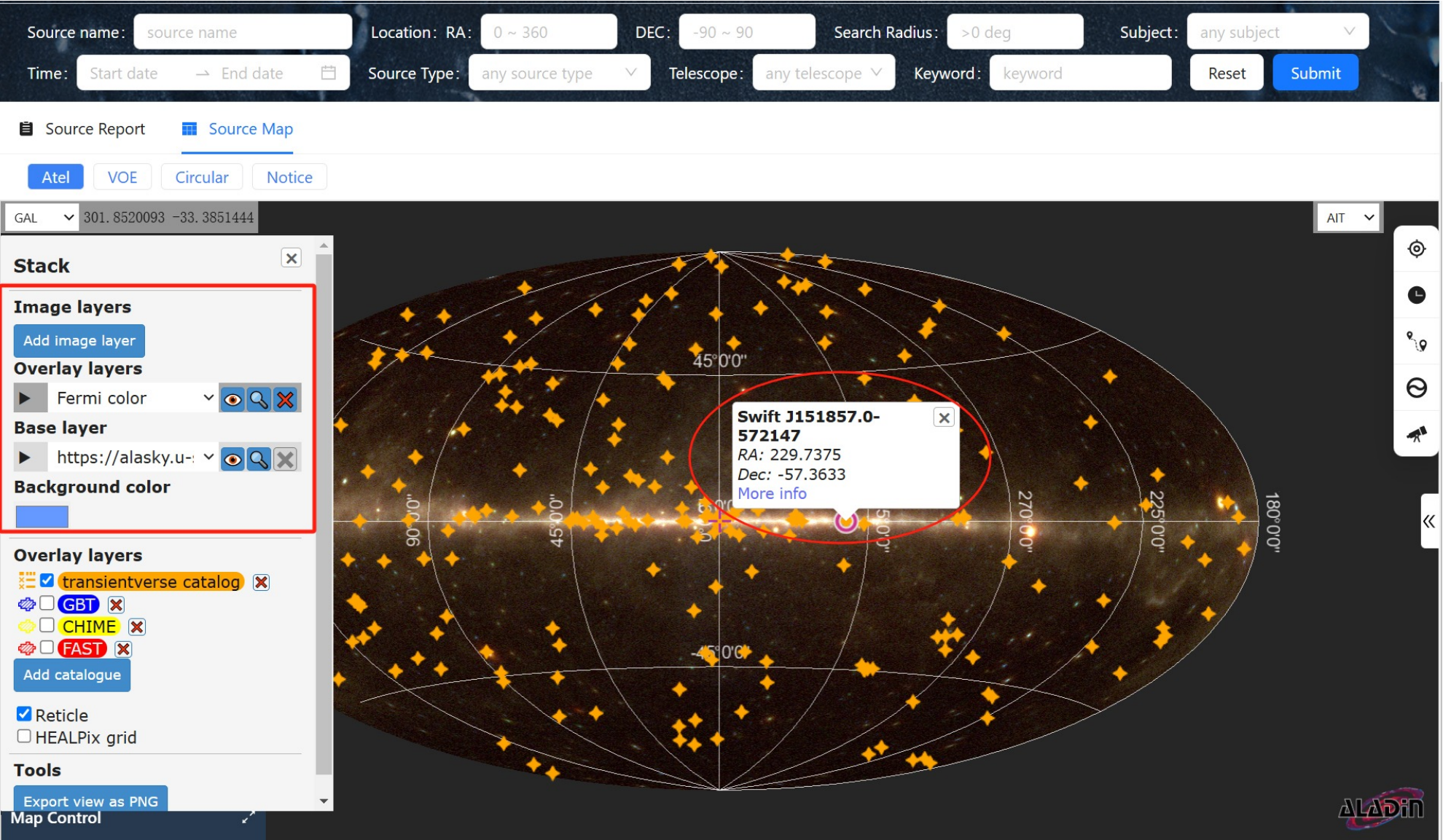}
    \caption{Interactive sky map with multi-wavelength overlays: This image shows an interactive sky map with multiple wavelength layers, including a Fermi color layer (highlighted in red). The map allows users to overlay different wavelength bands and visualize celestial sources, with an elliptical red frame marking a source that can be clicked for detailed information (such as basic data, historical alerts, and related papers from ADS). In the upper-left corner, users can select different coordinate systems; the lower-left corner displays the visible sky regions of different telescopes; and the upper-right corner offers different projection modes.}
    \label{fig:sky_map}
\end{figure*}

\begin{itemize} 
    \item \textbf{Textual Data Display}: Accessible by clicking the ``Report Title" in the structured data interface, this display presents the full text of the corresponding alert, preserving its original format. This ensures the integrity of the information and allows users to quickly browse the details.

    \item \textbf{Structured Data Display}:  
    Key information such as source name, coordinates, observation time, and telescope name, etc., is presented in a table (see Table \ref{tab:tab2}).
    For human-readable texts from Atel and Circulars, fields like RA, Dec,  RA\_hms, Dec\_hms, Loc\_err, Source\_Name, Report Subjects, Telescope, and Source\_Type are extracted via open-source LLMs, though the extraction accuracy is not guaranteed.
    Users can filter, analyze, and download selected reports in CSV format. 
    Clicking on the ``source name" reveals basic information retrieved from the SIMBAD\footnote{\url{https://simbad.u-strasbg.fr/simbad/sim-fbasic}} database, related reports, and papers from ADS. Clicking the ``Report Title" opens the original full alert message.
    
    \item \textbf{Interactive Sky Map Display}: As shown in Figure \ref{fig:sky_map}, this feature utilizes the Aladin\footnote{\url{https://aladin.cds.unistra.fr/}} astronomical visualization tool \citep{2022ASPC..532....7B}. It supports multiple coordinate projections (Aitoff, Mollweide, Orthographic, Mercator) and coordinate systems (Galactic and Equatorial). Moreover, it allows image overlays from visible light to radio wavelengths. Users can view source visualizations across different wavelengths, facilitating identification across bands. Additionally, to assist with potential follow-up observations, the platform provides telescope coverage maps. Users can select telescopes from the bottom-left corner, and their coverage areas are color-coded on the sky map.
\end{itemize}

The interlinking of these display methods enables seamless navigation. After applying filters, users can first view structured alert messages on the ``Source Report" page and then switch to the ``Source Map" to visualize the filtered sources on the sky map. Clicking either on a source name in the structured report or its corresponding point (yellow star) on the map will bring up detailed source information. Clicking the ``Report Title" in the structured message retrieves the full textual version of the alert message, enhancing analysis flexibility.

\subsubsection{Subscription of Transient Source Alerts}
To better serve astronomers, the TransientVerse platform offers a subscription feature, allowing users to select specific alert source types based on their interests and receive related alerts via email. The system automatically pushes notifications when new data is available, ensuring the timeliness and relevance of the information while reducing the need for users to frequently visit the platform. The subscription system is designed to be simple and intuitive, supporting customizable notification settings, enabling users to efficiently receive transient source alerts and facilitating follow-up multi-wavelength and multi-messenger observations.

\subsubsection{Quick Literature Survey for Transient Sources}\label{Quick Literature Survey for Transient Sources}
\begin{figure*}[ht]
\centering
\begin{minipage}{0.48\textwidth}
    \centering
    \includegraphics[width=\textwidth]{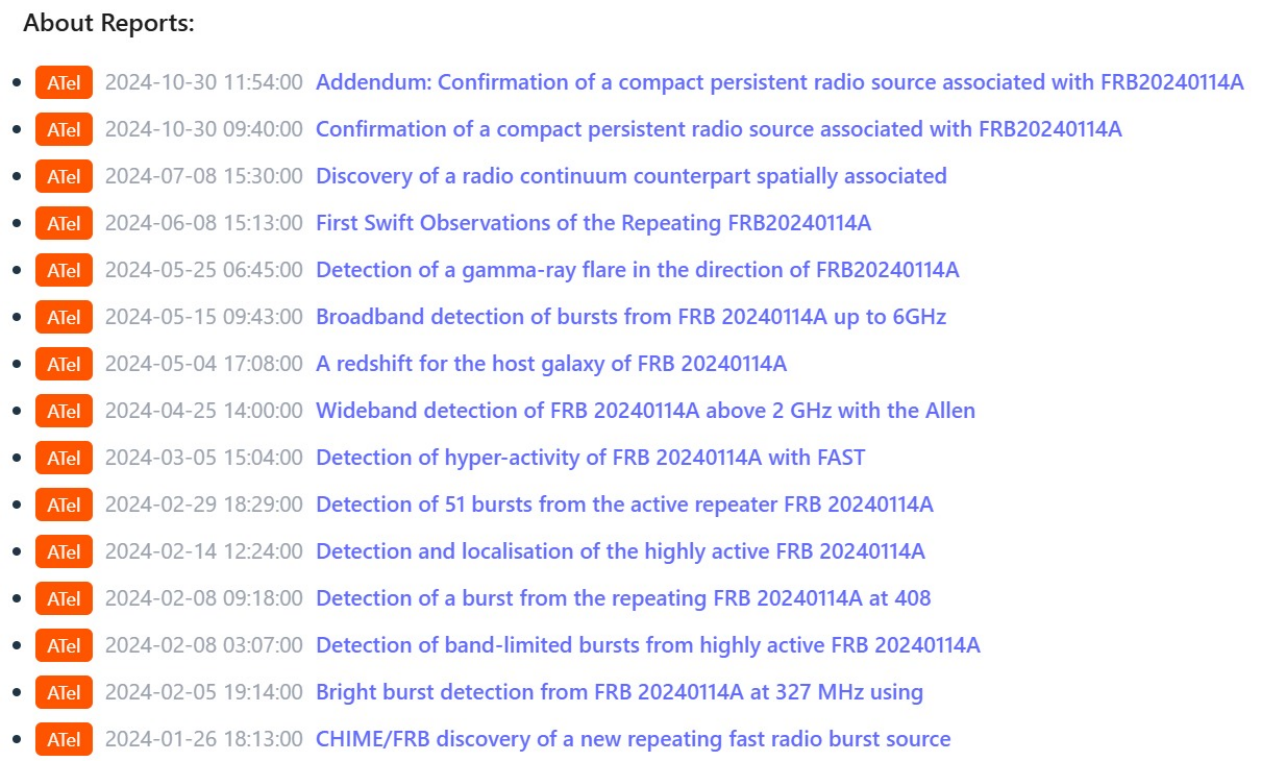}
\end{minipage}
\begin{minipage}{0.48\textwidth}
    \centering
    \includegraphics[width=\textwidth]{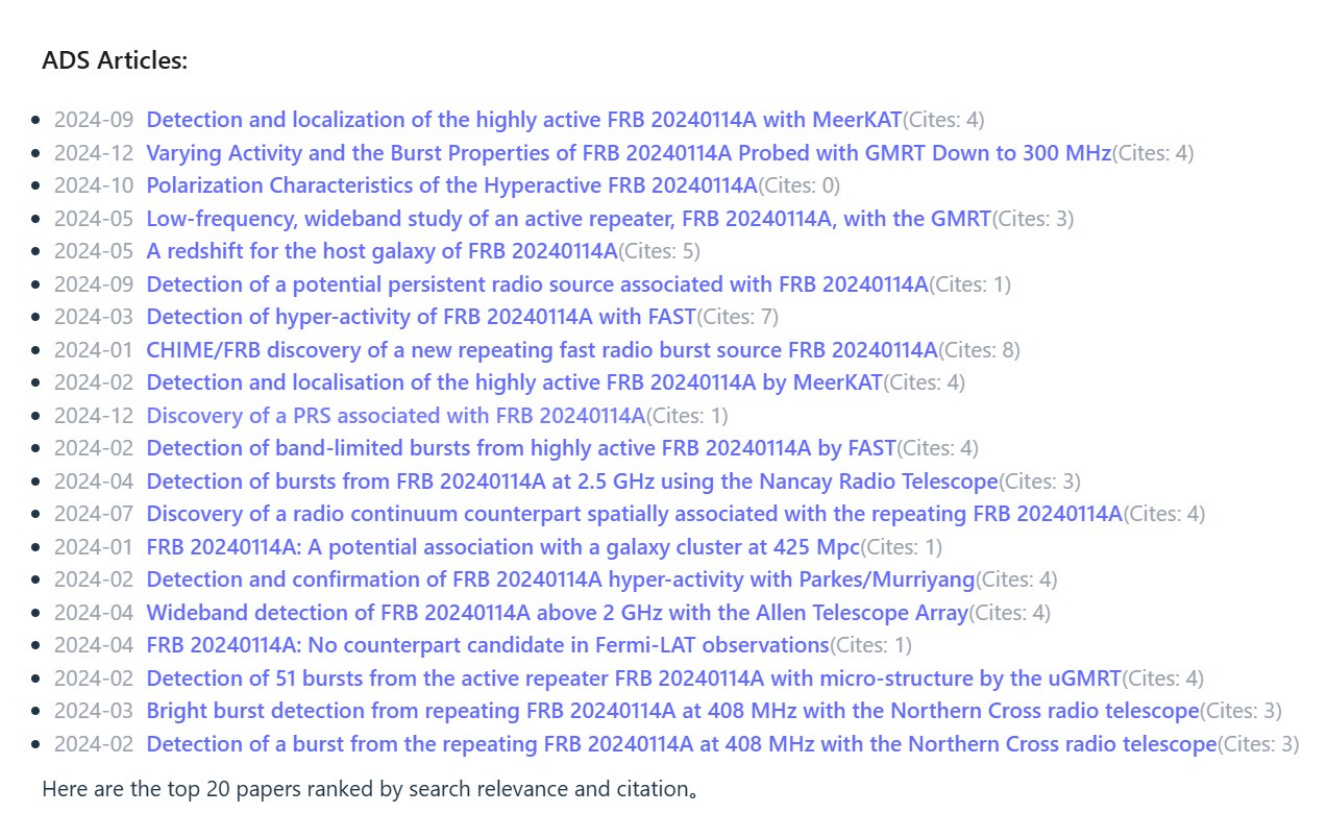}
    \put(-135,140){\textbf{(b)}}
\end{minipage}
\caption{A quick literature survey for the source FRB 20240114A, showing:(a) related alerts in chronological order, and (b) relevant papers from ADS.}
\label{fig:FRB20240114A}
\end{figure*}

When researchers receive new astronomical alerts, they often seek comprehensive information about the celestial sources mentioned, including their basic properties, observation history, and related research. On the TransientVerse platform, users can access detailed information about a celestial source through two methods. The first method is through the structured alert interface, as shown in Figure \ref{fig:transient_structed}, where clicking on a source name directs users to a page displaying three types of information: (a) basic data from the SIMBAD database, (b) historical observation alerts aggregated from ATel, GCN, and VOEvent, and (c) relevant literature retrieved from the Astrophysical Data System (ADS). The second method is through the Source Map, where users can click on a source point and select ``more info" to access the same three categories of information. An example of this process for the source FRB 20240114A is shown in Figure \ref{fig:FRB20240114A}.

Additionally, TransientVerse links the source name to our curated structured astronomical report database, aggregating historical observation reports about the source from sources like ATel, GCN, and VOEvent to provide comprehensive historical observation information.

To support in-depth research, the platform performs article searches within the Astrophysical Data System (ADS) using the extracted celestial names as keywords. The results are then sorted by relevance and citation count, showing only the top 20 papers. This feature ensures that researchers can efficiently access the most influential and valuable studies, thus improving research efficiency and workflow.

\subsubsection{Unit Conversion Toolkit}

To facilitate users in performing unit conversions while viewing transient source reports, the TransientVerse website provides unit conversion functions for coordinates, time, distance, and flux on the right side of the interface. The following are the specific functions:

\begin{itemize} 
    \item \textbf{Coordinate Unit Conversion}: Supports conversions between different coordinate systems (Equatorial, Galactic, and Ecliptic) and units (sexagesimal, degrees, and radians). Users can input coordinates and select the input and output celestial coordinate systems and units, and the system will automatically convert the coordinates to the output system and unit. 
    
    \item \textbf{Time Unit Conversion}: Supports conversion between time units such as UTC, MJD, and JD. After entering one unit, the system will automatically display the corresponding values in other units. When a user selects a telescope, the system will show the time in the time zone of that telescope. 
    
    \item \textbf{Distance Unit Conversion}: Supports conversions between units such as kilometers, astronomical units, parsecs, light years, and distance modulus (m-M). When a user enters a value in any of these units, the system will automatically convert it to all other units. 
    
    \item \textbf{Flux Unit Conversion}: Supports conversions between units such as Jy, $\rm AB~Mag$, $\rm W~m^{-2}~Hz^{-1}$, and $\rm erg~cm^{-2}~s^{-1}~Hz^{-1}$. When a user enters a value in any of these units, the system will automatically convert it to all other units. Furthermore, when the user inputs the equivalent pulse width (in milliseconds), the system will automatically calculate the energy density of the pulse in the unit ($\rm erg~m^{-2}~Hz^{-2}$). 
\end{itemize}

These unit conversion features support both \textbf{interactive unit conversion} and \textbf{batch unit conversion}. In batch conversion mode, after importing the file containing the units to be converted, users only need to select the corresponding unit in the file. The system will automatically display the converted values for all units in the output file, which users can then download directly.

\begin{figure*}[ht] 
    \centering 
    \includegraphics[width=1\textwidth]{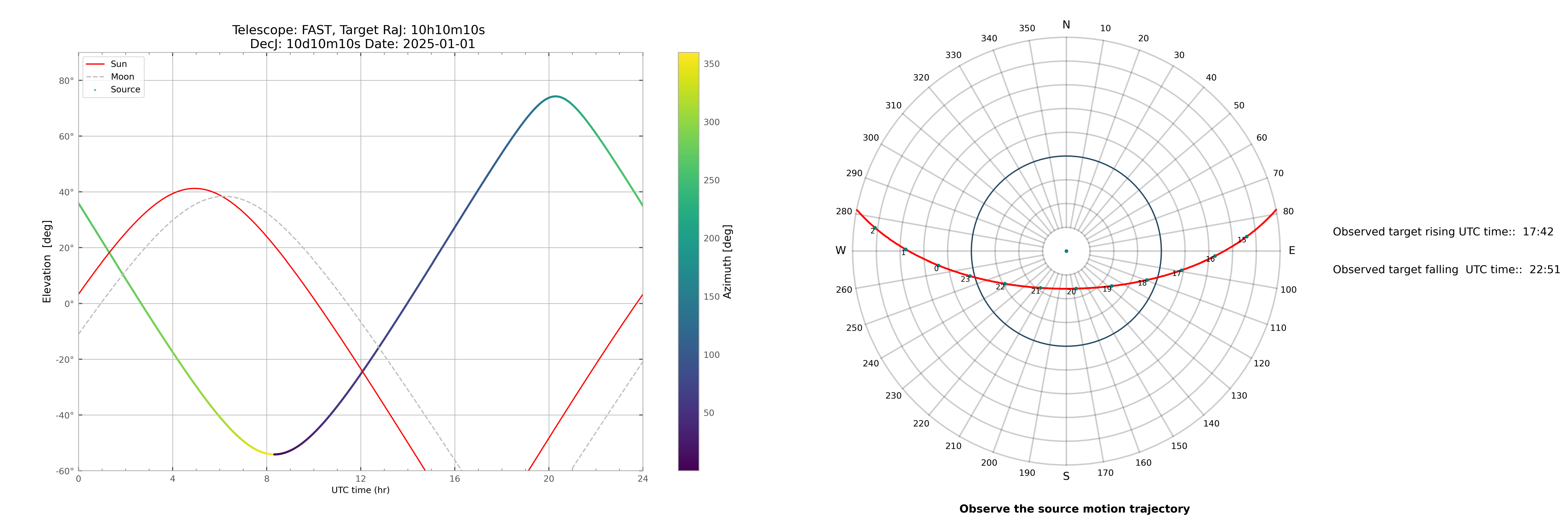} 
    \caption{Interactive observation planning tool. For instance, when setting the observation date to January 1, 2025, with the FAST telescope, source position at RaJ 10:10:10, DecJ 10:10:10, and a maximum zenith angle of 40$^{\circ}$, the tool shows the source’s elevation variation throughout the day. The left plot shows elevation versus UTC time with color-coded azimuths, moon and sun altitude (dashed and red lines, respectively). The right plot displays the target’s daily trajectory in polar coordinates, with azimuth and zenith angle, where the red curve represents the path and the green dot marks the position at the specified time. Precise rise and set times in UTC are shown on the right.}
\label{fig:tools5} 
\end{figure*}

In addition, to support users in conducting follow-up observations of queried sources, the tool provides the visibility time for certain sources from different telescopes. As shown in Figure \ref{fig:tools5}, when users input the source's coordinates, select a telescope, enter the observation time, and provide the telescope's maximum zenith angle, the system will display the source's elevation throughout the day for the selected telescope and provide the distribution of the zenith angle and azimuth at different times of the day. This figure illustrates an interactive tool for observation planning.

For instance, when the observation date is set to January 1, 2025, the telescope is configured to the Five-hundred-meter Aperture Spherical Telescope (FAST) \cite{2018IMMag..19..112L}, and the source coordinates are specified as RaJ 10:10:10 and DecJ 10:10:10, with a maximum zenith angle of 40$^{\circ}$, the system will display the variation in the source's elevation throughout the day. This information is visualized in two charts: the left chart shows the elevation variation as a function of UTC time, with different azimuths represented by distinct colors, as indicated by the color bar on the right. The dashed line represents the moon's altitude, while the red line indicates the sun's altitude. The right chart is a polar plot that illustrates the target's trajectory throughout the day, with azimuth shown along the angular direction and zenith angle represented as radial distance. The red line traces the target's motion, and the green dot marks its position at the specified time. In addition, the precise rise and set times of the source (in UTC) are displayed on the right.

\subsubsection{Follow-up Observation Alert Feature for FRBs Based on VOEvent Messages}

\begin{figure*}[htbp!] 
    \centering 
    \includegraphics[width=0.85\textwidth]{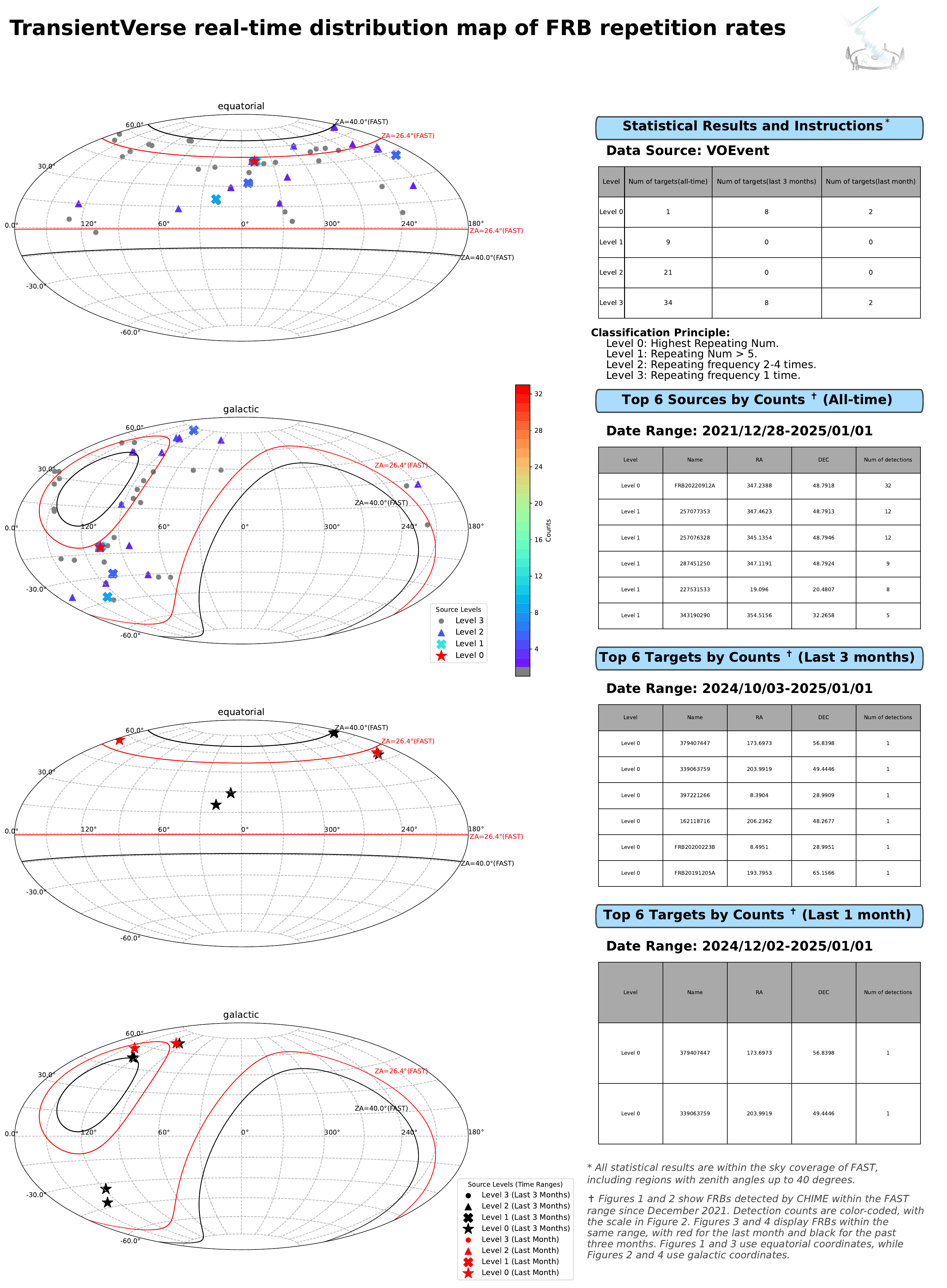} 
    \caption{Detection frequency statistics of repeating FRBs and high-frequency source information from the VOEvent system. The left panel includes four distribution plots: the top two show the detection frequency of repeating FRBs in equatorial and galactic coordinates, color-coded by frequency levels (based on the color bar in Figure 2). The bottom two show the detection frequencies of the same in the past three months (in black) and the past month (in red). Marker shapes and sizes represent repetition frequency levels, and red and black lines indicate the zenith angle constraints for the FAST telescope ($26.4^{\circ}$ and $40^{\circ}$). The right panel contains four tables: the first classifies sources by repetition frequency levels, while the other three list the top six highest-frequency sources for all time, the past three months, and the past month.} 
\label{fig:FRB} 
\end{figure*}

As shown in Figure \ref{fig:FRB}, this figure presents the detection frequency statistics of repeating FRBs and the high frequency source information from the VOEvent system. The left panel contains four detection frequency distribution plots for repeating FRBs with common elements. Solid red and black lines at $26.4^{\circ}$ and $40^{\circ}$ mark the zenith angle constraints for the FAST telescope. The detection frequencies of repeating FRBs are divided into four levels: Level 0 (highest), Level 1 ($\geq$ 5 detections), Level 2 (2 - 4 detections) and Level 3 (1 detection). Markers vary in shape and size, and the size is proportional to the detection frequency. The differences between the plots are in the data and color use. The top two plots show the detection frequency of all repeating FRBs in the equatorial and galactic coordinate systems, and color is used to distinguish repetition frequency levels based on the color bar in Figure 2. The bottom two plots display detection frequencies for the past three months and the past month, respectively, with red indicating the past month and black indicating the past three months, rather than frequency levels. The right panel contains four tables. The first classifies sources by repetition frequency levels, while the remaining three list the top six highest-frequency sources for all time, the past three months, and the past month. These visualizations provide an intuitive statistical overview, helping astronomers identify high-frequency erupting FRB sources for follow-up observations.

\section{Discussion and Conclusion}\label{sec:conclusion}
The TransientVerse has been designed with the objective of becoming the world's leading comprehensive platform for transient source alert management. It facilitates a range of services, including subscription management, retrospective queries, and rapid access to basic information, historical alert data, and related literature for sources referenced in the reports. In addition, the platform's visualization tools and user-friendly interface support real-time queries and personalized customization, significantly optimizing the research workflow. The modular design of message sources ensures the platform’s high scalability, allowing for flexible integration with external services and data sources, thus providing strong support for multi-wavelength and multi-messenger astronomy research.

In the field of transient source astronomy, another platform similar to \texttt{TransientVerse} is \texttt{Astro-COLIBRI} \cite{2021ApJS..256....5R}\footnote{\url{https://astro-colibri.org/}}. Although \texttt{Astro-COLIBRI} shares many features with \texttt{TransientVerse}, its design philosophy and application focus seem to differ in several ways. 
\texttt{Astro-COLIBRI} appears to place a greater emphasis on the detailed analysis of individual alert events. The event detail pages feature a discussion area where astronomers can exchange insights and share opinions on specific astronomical events, potentially fostering academic discourse. Additionally, the platform offers a mobile app, which further extends its accessibility and usability.
In contrast, \texttt{TransientVerse} emphasizes the in-depth exploration of astronomical sources associated with alert events. By leveraging LLMs and custom parsers, the platform extracts source names from alert messages, retrieves essential source information from SIMBAD, searches related alert data within its database, and gathers relevant literature from ADS. This facilitates a rapid and comprehensive understanding of the sources involved in each event. Moreover, the platform features an observational assistant tool, which aids astronomers in determining the necessity of follow-up observations, thereby enabling multi-wavelength and multi-messenger collaborative efforts that drive progress in astrophysical discoveries. 

TransientVerse platform presents several remarkable features and advantages as follows:
\begin{itemize}
    \item \textbf{Integration and Structuring of Alert Messages:}  
    The platform integrates alert messages from various transient alert systems and employs open-source LLMs and custom parsers to convert textual alerts into structured data. This improves data consistency and query efficiency, laying the foundation for further analysis and visualization. Its modular design also supports the addition of new alert sources and telescope systems.
    \item \textbf{Historical Data Retrieval and Query:}  
    Users can flexibly query historical alert messages based on parameters such as time range and source type. The platform also supports downloading structured data for subsequent scientific analysis.
    \item \textbf{Sky Map Visualization:}  
    The platform provides a sky map-based visualization of transient sources, supporting multiple coordinate systems and projection modes. Users can overlay specific telescope visibility regions to quickly determine whether a target is within the observable range, thereby improving observation planning efficiency.
    \item \textbf{Source Investigation and Literature Support:}  
    By clicking on a transient source's name, users can access detailed information, including historical alert records and related academic literature. The platform leverages the ADS system to perform precise searches for the literature associated with the source, helping researchers to obtain reference materials efficiently.
    \item \textbf{Observation Tools and Optimal Observation Time Recommendation:}  
    The platform offers practical tools for tasks such as flux, time, distance, and coordinate unit conversions. Through the ``Observation Assistant" module, users can obtain intelligent recommendations for optimal observation times based on observation dates and telescope specifications, maximizing the utility of limited observation windows.
\end{itemize}

The platform will continue to expand its data ecosystem by incorporating additional telescopes and alert responses, enhancing both the timeliness and coverage of event detection. This will not only enrich the available data resources but also further improve the platform's response speed and analytical capabilities for transient events. In parallel, the platform plans to introduce a multi-model matching feature for FRBs, enabling more sophisticated event analysis. By integrating various theoretical models and observational data, the platform will be able to achieve more accurate event identification and prediction, advancing the study and understanding of FRB mechanisms. This expansion will not only meet the needs of current users but also pave the way for future developments in multi-messenger astronomy.

The TransientVerse development team welcomes feedback from the astronomy community to further improve the platform. Please feel free to contact us via the authors' email. \footnote{\url{https://transientverse.zero2x.org/about}}

\section*{Acknowledgments}
This work is supported by National Natural Science Foundation of China (NSFC) Programs (No. 11988101, U1731238); by CAS International Partnership Program (No. 114-A11KYSB20160008); by CAS Strategic Priority Research Program (No. XDB23000000); and the National Key R\&D Program of China (No. 2017YFA0402600, 2023YFC2206403, 2022ZD0115305); and the National SKA Program of China (No. 2020SKA0120200, 2022SKA0130100). 
D.L. is a New Cornerstone investigator. 
P.W. acknowledges support from the CAS Youth Interdisciplinary Team, the Youth Innovation Promotion Association CAS (id. 2021055), and the Cultivation Project for FAST Scientific Payoff and Research Achievement of CAMS-CAS.
J.H.F. and H.W. acknowledge support from the National Key R\&D Program of China (No. 2023YFC2206403). 
H.X.C. acknowledges support from the  National Key R\&D Program of China (No. 2022ZD0115305). 
This document was prepared using the collaborative tool Overleaf available at: \url{https://www.overleaf.com/}.

\bibliography{sample631}{}
\bibliographystyle{aasjournal}

\end{document}